\documentclass[aps,prl,10pt,twocolumn,superscriptaddress,showpacs,preprintnumbers,nofootinbib]{revtex4-1}
\usepackage{titlesec}
\usepackage{hyperref}
\usepackage{graphicx}
\usepackage{amsfonts,amsmath,amssymb,bm,bbm}
\usepackage{color}

\newcommand{\mm}[1]{\mathrm{#1}}
\renewcommand{\vec}[1]{\mathbf{#1}}

\renewcommand{\Im}{\mbox{Im}}
\newcommand{\D}{{\mathrm{d}}}

\hypersetup{
    pdfnewwindow=true,      
    colorlinks=true,       
    linkcolor=blue,          
    citecolor=blue,        
    filecolor=blue,      
    urlcolor=blue        
}

\titleformat{\section}[runin]
  {\normalfont\itshape}{\thesection}{1em}{}[.---]
\titlespacing*{\section}{0.5cm}{2em}{1em}
\titleformat{\subsection}[runin]
  {\normalfont\itshape}{\thesubsection}{1em}{}[---]
\titlespacing*{\subsection}{1cm}{2em}{1em}

\begin{document}

\preprint{SLAC-PUB-17174}
\preprint{MITP/17-090}

\title{Diffuse axion-like particle searches}
\author{Hendrik Vogel}
\email{hvogel@slac.stanford.edu}
\affiliation{SLAC National Accelerator Laboratory, Stanford University, Menlo Park, 94025, USA}

\author{Ranjan Laha}
\email{ranjalah@uni-mainz.de}
\affiliation{PRISMA Cluster of Excellence and
             Mainz Institute for Theoretical Physics,
             Johannes Gutenberg-Universit\"{a}t Mainz, 55099 Mainz, Germany}

\author{Manuel Meyer}
\email{mameyer@stanford.edu}
\affiliation{SLAC National Accelerator Laboratory, Stanford University, Menlo Park, 94025, USA}
\affiliation{W. W. Hansen Experimental Physics Laboratory \& Kavli Institute for Particle Astrophysics and Cosmology,
Stanford University, Stanford 94305, USA}

\date{\today}

\begin{abstract}
We propose a new method to search for axion-like particles (ALPs) based on the gamma-rays produced concomitant with high-energy astrophysical neutrinos.
The existence of high-energy neutrinos implies production of gamma-rays in the same sources. Photons can convert into ALPs in the sources' magnetic fields, and will travel as ALPs through extragalactic space. Back-conversion in the Milky Way's magnetic field leads to a diffuse anisotropic high-energy photon flux that existing and upcoming gamma-ray detectors, like HAWC, CTA, and LHAASO can detect. This method probes unexplored ALP parameter space, with LHAASO being realistically sensitive to couplings above $10^{-11}\, \rm{GeV^{-1}}$ and masses up to $3\times 10^{-6} \, \rm{eV}$ in ten years. Our technique also explores viable ALP dark matter parameter space. 
\end{abstract}

\maketitle


IceCube's detection of diffuse high-energy ($> 10$ TeV) astrophysical neutrinos~\cite{Aartsen:2013bka,Aartsen:2017mau,Aartsen:2016xlq} presents new challenges and exciting opportunities for particle astrophysics. These opportunities include new physics searches\,\cite{Feldstein:2013kka,Murase:2015gea,Esmaili:2015xpa,Bhattacharya:2016tma,Dev:2016qbd,Bhattacharya:2017jaw} and new IceCube signals\,\cite{Kistler:2016ask,Li:2016kra}. IceCube's high-energy neutrinos are not produced in our atmosphere~\cite{Aartsen:2015nss,Bhattacharya:2016jce,Bhattacharya:2015jpa,Gauld:2015kvh,Gauld:2015yia,Benzke:2017yjn,Garzelli:2016xmx,Garzelli:2015psa,Halzen:2016pwl,Halzen:2016thi,Laha:2016dri,Goncalves:2017lvq,Laha:2013eev}, and no correlation with the Milky Way (MW)~\cite{Aartsen:2017ujz} or any astrophysical source has been found~\cite{Aartsen:2016tpb}. Most likely, multiple sources contribute to this seemingly diffuse extragalactic flux\,\cite{Feyereisen:2017fnk}.
\begin{figure}[h!]
        \includegraphics[width=0.48\textwidth]{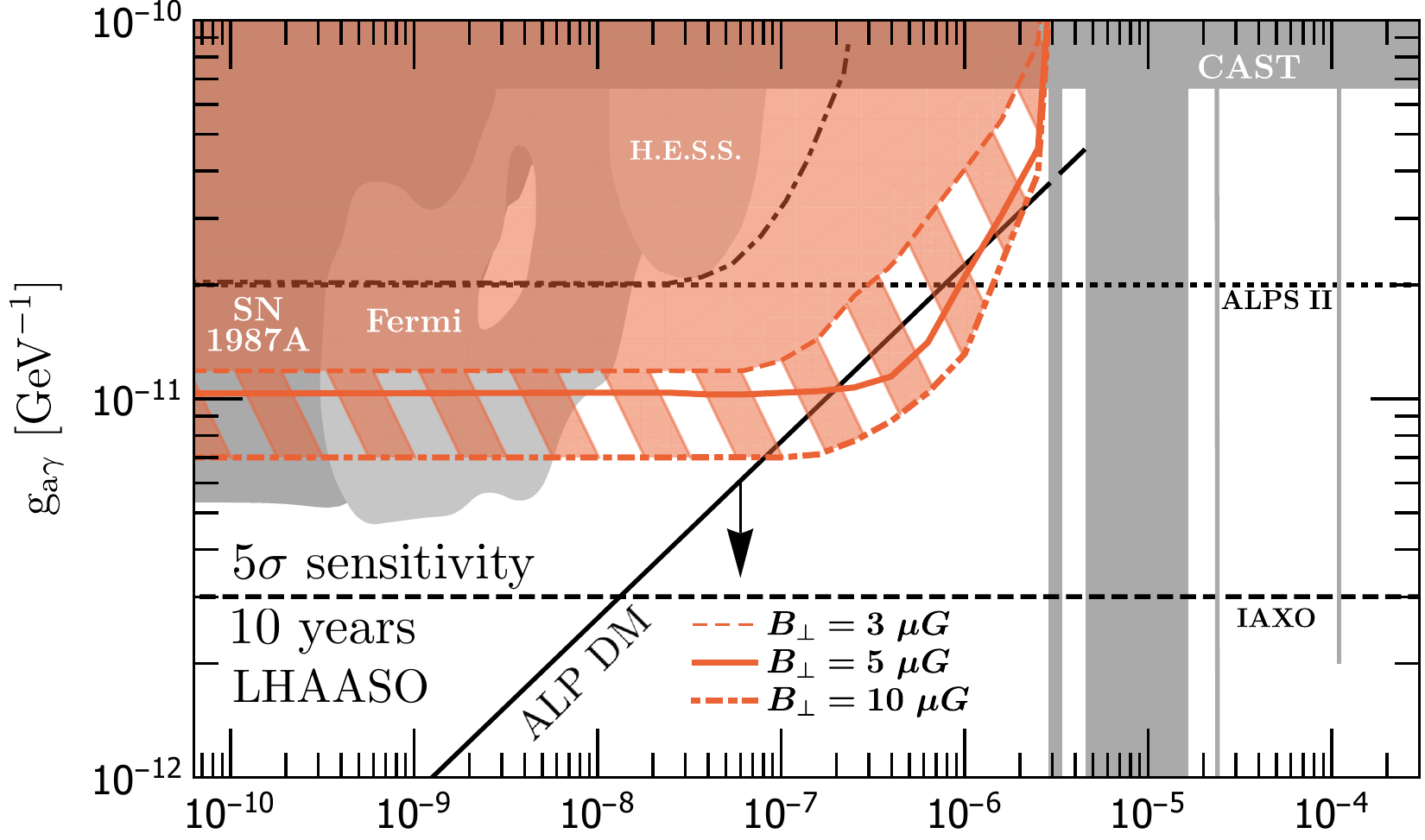}\\
         \includegraphics[width=0.48\textwidth]{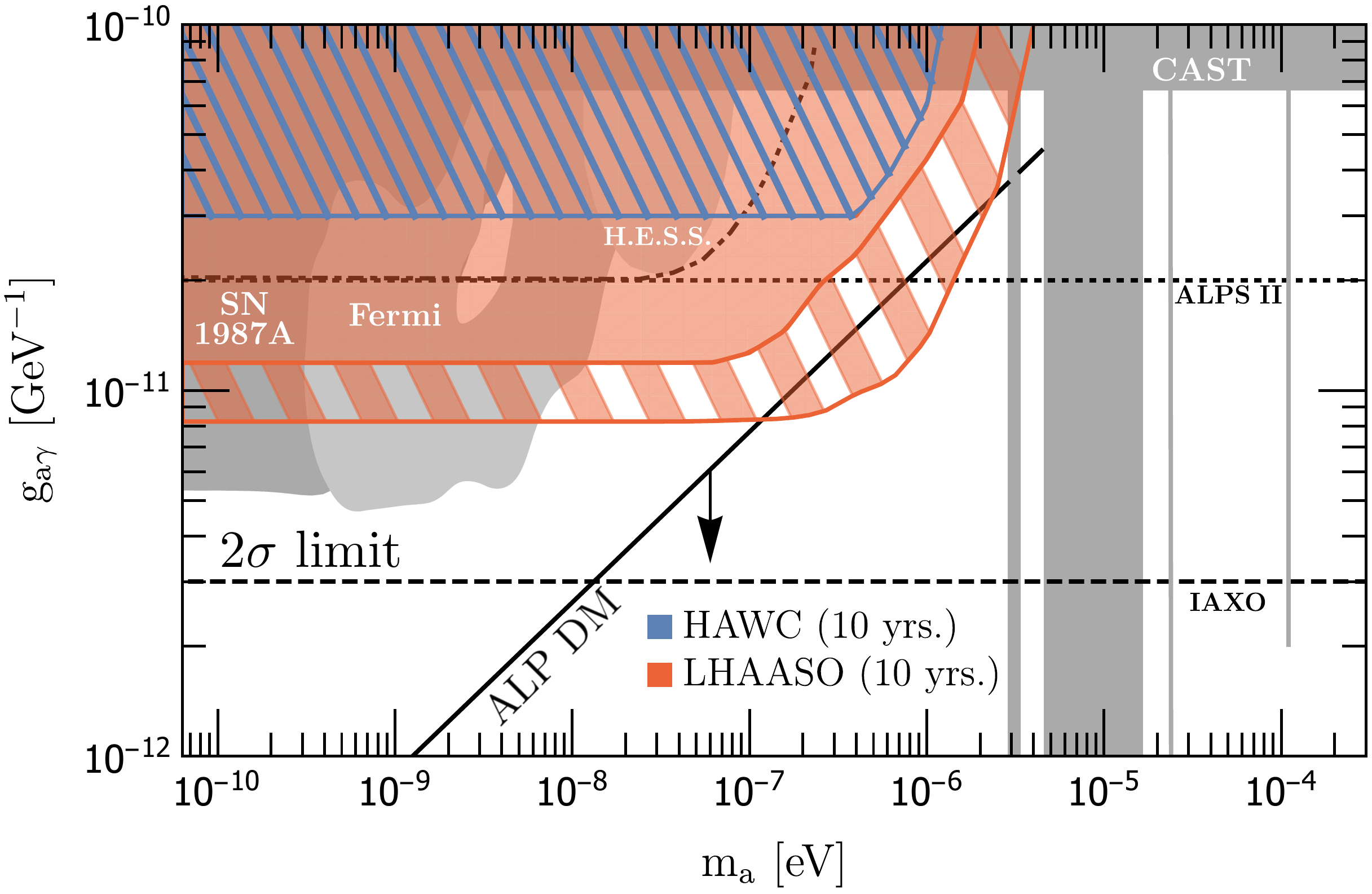}
        
  \caption{{\emph{Top:}} LHAASO's sensitivity (red) to a diffuse ALP flux for the realistic scenario. The hatched region demonstrates the variation due to the source $\vec{B}$-fields ($B_\perp=3,5,10\, \mu$G). We assume a coherence length of 1 kpc. The shaded regions are limits from CAST\,\cite{Anastassopoulos:2017ftl}, various haloscopes\,\cite{McAllister:2017lkb,Brubaker:2016ktl,Asztalos:2009yp,Hagmann:1990tj,Wuensch:1989sa,DePanfilis:1987dk}, and astrophysics~\cite{TheFermi-LAT:2016zue,Abramowski:2013oea,Payez:2014xsa}.  We also show future sensitivities of CTA\,\cite{Meyer:2014gta}, ALPS-II\,\cite{Bahre:2013ywa}, and IAXO\,\cite{Irastorza:2013kda}.  The solid black line (``ALP DM") shows the parameter space below which ALPs can act as dark matter\,\cite{Arias:2012az}. {\emph{Bottom:}} Limits from HAWC (blue) and LHAASO (red). Hatched regions show the uncertainty from our assumptions: lower (upper) boundary is for the realistic (conservative) scenario. The upper boundary is not visible for HAWC.}\label{fig:MainResult}
\end{figure}
Extragalactic neutrinos are produced through hadronic processes of cosmic-rays with ambient gas ($p-p$) or radiation fields ($p-\gamma$). These interactions create concomitant high-energy photons (10 TeV -- 10 PeV), which are efficiently absorbed with a mean free path $\lesssim$ 100 Mpc~\cite{Gould:1966pza,Ruffini:2015oha,Venters:2010bq}. No significant correlation between photons and neutrinos has been observed.
 
Axion-like particles (ALPs) can significantly enhance the photon-flux at Earth through the light-shining-through-the-wall mechanism~\cite{Anselm:1986gz,Sikivie:1983ip,VanBibber:1987rq,DeAngelis:2007dqd,Simet:2007sa,Mirizzi:2009aj,SanchezConde:2009wu,DeAngelis:2011id,Horns:2012kw,Meyer:2014epa,Galanti:2015rda,Montanino:2017ara}.
    ALPs couple to photons via the Lagrangian $g_{a \gamma} \, a \,\vec{E}\cdot \vec{B}$, where $g_{a \gamma}$ is the coupling strength, $a$ the ALP, $\vec{E}$ the photon's electric field, and $\vec{B}$ an external magnetic field.  This term mediates photon-ALP oscillations with a maximal probability $P_{a\gamma}\propto (g_{a\gamma}B\,l)^2$ in a homogeneous $\vec{B}$-field of length $l$.  While laboratory-based searches for ALPs use strong magnetic fields\,\cite{Anastassopoulos:2017ftl,Ehret:2009sq, Bahre:2013ywa,Irastorza:2013kda}, the vast extent of the weak galaxy and cluster $\vec{B}$-fields\,\cite{Fletcher:2011fn,Carilli:2001hj,Beck:2016,2017ARA&A..55..111H} leads to large oscillation probabilities~\cite{Mirizzi:2007hr,Hooper:2007bq,Simet:2007sa,Hochmuth:2007hk,DeAngelis:2007wiw,Chelouche:2008ta,Jimenez:2011pg,Conlon:2013txa,Fairbairn:2013gsa,Conlon:2014xsa,Meyer:2016xve}. ALPs travel unhindered through extragalactic space (assuming small $\vec{B}$-fields) and some will encounter the MW.  In the MW's magnetic field\,\cite{Jansson:2012pc,Jansson:2012rt,Unger:2017kfh}, these ALPs will partially back-convert into photons.

Here, we point out that these diffuse gamma-rays are detectable in current and near-future experiments like the High Altitude Water Cherenkov (HAWC)~\cite{Abeysekara:2017hyn}, the Cherenkov Telescope Array (CTA)~\cite{Acharya:2017ttl}, and the Large High Altitude Air Shower Observatory (LHAASO)\,\cite{DiSciascio:2016rgi,Vernetto:2016gro,Vernetto:2015urm,Liu:2016kro,Cui:2014bda}.  Due to photon-ALP conversions, up to one-third of the initially produced photons survive this journey~\cite{Grossman:2002by}, which provides a detectable flux with little background. This is the {\it first} search for new physics using {\it only} these high-energy photons and we show that the prospects are extremely encouraging.

 All high-energy astrophysical searches for ALPs concentrate on one particular source class\,\cite{Conlon:2017qcw,Gondolo:2008dd,Friedland:2012hj,Viaux:2013lha,Meyer:2016wrm,Fischer:2016cyd,Payez:2014xsa,Ayala:2014pea,Abramowski:2013oea,TheFermi-LAT:2016zue,Isern:2008nt}. 
    Instead, our search technique relies on the integrated diffuse ALP flux by all neutrino sources. Since the initial photons are efficiently absorbed, a detection of such non-Galactic photons is a striking signature of ALPs that is hard to explain with astrophysical variability.  Our technique is also applicable when individual sources are undetected.

A background photon-flux to this search is produced by cosmic-ray interactions with the interstellar material in the MW itself.  This background is typically confined close to the Galactic disc where gas is abundant. On the other hand, these Galactic photons can convert to ALPs in the $\vec{B}$-field of the MW, and thus lead to spectral distortions similar to Refs.~\cite{Mirizzi:2007hr,Hooper:2007bq,Hochmuth:2007hk,DeAngelis:2007wiw,Chelouche:2008ta,SanchezConde:2009wu,Jimenez:2011pg,Meyer:2016xve}, which also can be used to search for ALPs.

The strength of this new method is illustrated in Fig.~\ref{fig:MainResult}. We show  projected $5\sigma$ sensitivities (top) and $2\sigma$ limits (bottom) to the gamma-rays associated with the diffuse ALP-flux as a function of $g_{a\gamma}$ and the ALP mass $m_a$.  Our method will likely probe parameter space where ALPs can contribute to dark matter (below the black line ``ALP DM'')~\cite{Arias:2012az}.

 Since our predicted fluxes depend on unknown source parameters, we consider a realistic and a conservative scenario, for which we show the assumptions in Tab.~\ref{tab:scenarios}. The first parameter is the sources' magnetic field perpendicular to the line-of-sight, $B_\perp$, for which a realistic estimate for an ensemble of generic sources is $B_\perp =5 \, \mu \mm{G}$~\cite{Fletcher:2011fn}. The other uncertainties are the source distribution with redshift, and the neutrino spectrum.

\begin{table}[b]

\caption{Summary of assumptions for the realistic and conservative scenarios that we consider. We vary $B_\perp$, source-evolution models, and astrophysical neutrino spectra.}

\begin{ruledtabular}

\begin{tabular}{lccc}
Scenario & $B_\perp$ [$\mu$G] & Source evolution & Neutrino spectrum \\ 
\hline
Realistic & 5 & Y\"uksel~\cite{Yuksel:2008cu} & Kopper~\cite{Aartsen:2017mau}\\
Conservative & 3 & HS~\cite{Hernquist:2002rg} & Niederhausen~\cite{Aartsen:2017mau}\\

\end{tabular}

\end{ruledtabular}
\label{tab:scenarios}
\end{table}

{\itshape High-energy photons and photon-ALP mixing.---} The normalization and shape of IceCube's neutrino flux depends on the energy range that is considered~\cite{Aartsen:2013bka,Aartsen:2013jdh,Aartsen:2017mau}. We focus on two fluxes taken from Kopper's and Niederhausen's contribution to Ref.~\cite{Aartsen:2017mau}, which we parametrize with
\begin{equation}
\phi(E_\nu) \propto \left[\left(\frac{E_\nu}{E_b}\right)^2+\left(\frac{E_\nu}{E_b}\right)^{2\alpha}\right]^{-\frac{1}{2}}
\label{eq:neutrino spectrum}
\end{equation}
where $E_\nu$ is the neutrino energy, and $E_b= 40\, (12) \,\mm{TeV}$ is the break energy for the realistic (conservative) scenario, and $\alpha=2.92 \, (2.48)$ the spectral index. The per-flavor normalization is $2.46 \,(1.57)\times 10^{-18} \, \mm{GeV^{-1}cm^{-2}s^{-1}sr^{-1}}$ at $E_\nu = 100 \, \mm{TeV}$.

In the following, we assume that these neutrinos are produced by $p-\gamma$, since $p-p$ is more strongly constrained by gamma-ray measurements~\cite{Kistler:2015oae,Ando:2015bva}. The energy of the photon is then $E_\gamma \approx 2 \, E_\nu$.  Each interaction produces two photons per three neutrinos\,\cite{Kelner:2008ke,Kistler:2015ywn}.  These relations permit us to construct the source photon spectrum using Eqn.\,\ref{eq:neutrino spectrum}.

If ALPs exist, these photons will partially oscillate into ALPs in the source's  $\vec{B}$-field\,\cite{Fletcher:2011fn,Carilli:2001hj,Beck:2016,2017ARA&A..55..111H}.  The probability depends on the interstellar photon and electron density, and $\vec{B}$.  These parameters are not precisely known since we lack information on the neutrinos' sources.  They will also show variability between sources and evolve with redshift. However, our approach does not rely on a single source but on ensembles and mean environmental parameters. We therefore vary these parameters in motivated ranges and assume generic source evolution models.

Our mean $z{=}0$ $\vec{B}$-field strength is inspired by Ref.~\cite{Fletcher:2011fn}, which provides a distribution of measured field strength for the regular component perpendicular to the line-of-sight $B_\perp$.  We use the domain model~\cite{DeAngelis:2007dqd,Mirizzi:2009aj,DeAngelis:2011id}, which splits the magnetic field into domains with a fixed length. We choose this length at $z{=}0$ to be $l=1\,\mm{kpc}$. In each domain the $\vec{B}$-field strength and direction are constant and follow a probability distribution. We model each component of $\vec{B}$ as a Gaussian distribution with zero mean and variance $2B_\perp^2/3$, so that $\left\langle|\vec{B}_\perp|\right\rangle\approx B_\perp$.  We assume that the magnetic field extends up to $7\, \mm{kpc}$ for $z{=}0$ galaxies. The redshift evolution of these parameters and the electron density is taken from Ref.~\cite{Schober:2016ebm}.  For the interstellar photon background we use Ref.~\cite{Schober:2014dda}.

We apply the formalism presented in Ref.~\cite{Kartavtsev:2016doq} to obtain the mean ALP-flux from the source. Note that this formalism entails an average over the $\vec{B}$-field probability distribution. Here, this average is a true sample mean over a collection of galaxies. We do not suffer from the usual issue of what is meant by such an averaging when applied to individual sources. Equipartition implies that the maximum ALP-flux from a sample of sources is $1/3$ of the photon-flux.

The remaining photons are either attenuated in the source or in the extragalactic photon background\,\cite{Stecker:2016fsg,Franceschini:2017iwq}, while ALPs travel unimpeded.  We assume that the extragalactic magnetic field does not saturate its upper bounds~\cite{Ade:2015cva,Blasi:1999hu}, so that photon-ALP conversion is inefficient in extragalactic space~\cite{Kartavtsev:2016doq}.

When the ALPs reach the MW, they back-convert in the regular Galactic $\vec{B}$-field. We model the regular $\vec{B}$-field using the state-of-the-art fit~\cite{Jansson:2012pc,Jansson:2012rt,Unger:2017kfh}. Their model requires an off-the-plane component, with $<10\%$ normalization uncertainty.

 ALP-photon conversion can be computed with the density-matrix formalism~\cite{Raffelt:1987im}, which we modify to include absorption and local gamma-ray production
\begin{equation}\label{eq:propagation}
i\frac{\D}{\D z} \bm{\rho} = \left[\mathbf{H}_{\rm dis}, \bm{\rho}\right]-\frac{i}{2}\left\{\mathbf{H}_{\rm abs}, \bm{\rho}\right\}+i\mathbf{Q}\,,
\end{equation}
where $\bm{\rho}$ is the density matrix which contains particle densities and coherence, $\mathbf{H}_{\rm dis}$ and $\mathbf{H}_{\rm abs}$ are the dispersive and absorptive Hamiltonian respectively~\cite{Kartavtsev:2016doq}.  The source term for Galactic $p-p$ gamma-ray production  is $\mathbf{Q}$.  We computed its contribution using the cross sections given in Ref.~\cite{Kelner:2006tc}, assuming a homogeneous and isotropic cosmic-ray spectrum~\cite{Gaisser:2013bla}, and different gas models~\cite{Misiriotis:2006qq, Ahlers:2013xia}. The normalization of $\mathbf{Q}$ can be found by comparing Eq.~\ref{eq:propagation} to Eq.~1.23 in Ref.~\cite{rybicki1979radiative} for $g_{a\gamma}=0$.

Since our knowledge about the Galactic $\vec{B}$-field is much more precise compared to those in other galaxies, no averaging is required and the density-matrix equation can be solved directly.  Without a statistical distribution of magnetic fields, in some directions the full ALP-flux is converted into photons.  Note that the photon-flux is attenuated in the Galaxy as well, which we model using Ref.~\cite{Vernetto:2016alq}.

We include all dispersive terms~\cite{Kartavtsev:2016doq}. The oscillation frequency reads
\begin{align}
\Delta_{\rm osc} &= \left[(\Delta_\parallel -\Delta_a -\frac{i}{2}\Gamma)^2 +(2\Delta_{a\gamma})^2\right]^{\frac{1}{2}}\,,\\
\Delta_\parallel &= 2\Delta_{\rm B}+\Delta_{\gamma \gamma}+\Delta_{\rm pl} \,,
\end{align}
where $\Delta_{\rm B} \propto B^2$ is the magnetic birefringence,  $\Delta_{\gamma\gamma}$ the photon-photon dispersion~\cite{Toll:1952,Dobrynina:2014qba}, $\Delta_{\rm pl}$  the plasma frequency, and $\Gamma$ the absorption coefficient. The ALP mass is contained in $\Delta_a=- m_a^2/(2E_\gamma)$, and photon-ALP mixing is given by $\Delta_{a\gamma}\propto g_{a\gamma}B$. We compute photon-photon dispersion on interstellar radiation fields following Ref.~\cite{Dobrynina:2014qba}. At low energies, the ALP mass tends to suppress oscillations. The advantage of using high-energy photons in this method is that $m_a$ is relatively less important compared to existing astrophysical searches. It starts to dominate only around $m_a \sim 2\times 10^{-7} \, \mm{eV}$.

At large energies, $\Delta_{\rm B}$ tends to dominate $\Delta_\parallel$ and suppresses the high-energy part of the ALP spectrum. Interestingly, also $\Delta_{\gamma\gamma}$ turns out to be important in both sources and MW. For TeV photons, pair creation is only possible on a fraction of the interstellar radiation field and $\Delta_{\gamma \gamma}$ is positive~\cite{Dobrynina:2014qba}. In this regime, $\Delta_{\gamma \gamma}$ dominates the dispersion when $B$ is small in some domain of the source, or in, e.g., interarm regions of the MW.  

At larger energies, more background photons can undergo pair creation, which turns $\Delta_{\gamma \gamma}$ negative. This effect is enhanced for distant galaxies by the redshift evolution of the cosmic microwave background.  If $\Delta_{\gamma \gamma}$ is negative, it amplifies oscillations via a partial cancellation with $\Delta_\mm{B}$, or suppresses oscillations when $|\Delta_{\gamma\gamma}| \gg |\Delta_{\mm{B}}|$.

Note that when $\Delta_\parallel$ changes sign, the physical interpretation of the eigenvalues $E_{2,3}=\frac{1}{2}(\Delta_\parallel +\Delta_a \pm \Delta_{\rm osc} -\frac{i}{2}\Gamma)$ is switched. If $\Delta_\parallel - \Delta_a$ is positive, then the imaginary part of $\Delta_{\rm osc}$ is negative and $\Im \, E_2 \approx -\frac{i}{2}\Gamma$. In this case, $E_2$ can be interpreted as photon-like, whereas $E_3$ is ALP-like. If $\Delta_\parallel -\Delta_a$ turns negative, $E_2$ and $E_3$ interchange their imaginary part and $E_2$ is now ALP-like. This identification has to be taken into account in the transfer matrix formalism like in Ref.~\cite{Kartavtsev:2016doq}, unless the equations are chosen to be symmetric in $E_{2,3}$ as done in Ref.~\cite{DeAngelis:2007dqd}.

\begin{figure}
       \includegraphics[width=0.5\textwidth]{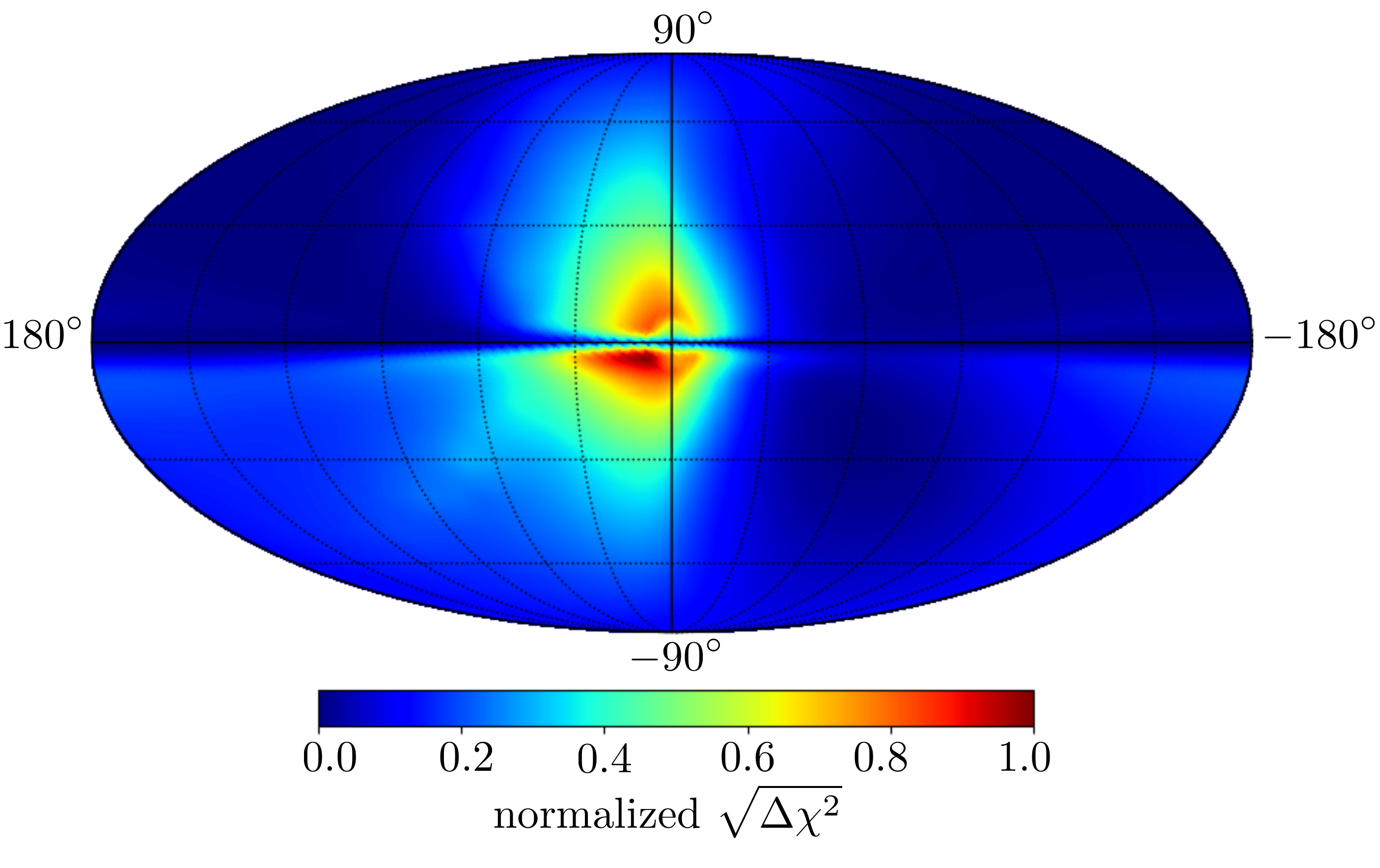}
  \caption{Relative sensitivity to the photon-flux from a diffuse-ALP flux for a LHAASO-like experiment.}\label{fig:GalaxyMap}
\end{figure}

{\itshape Results.---}The distribution of IceCube's high-energy neutrinos is isotropic in the sky.  The resultant isotropic ALP-flux incident on the MW will convert to high-energy photons in the presence of the regular Galactic $\vec{B}$-field\,\cite{Jansson:2012pc,Jansson:2012rt,Unger:2017kfh}, and its anisotropy will be imprinted on the resulting photon-flux (see Fig.~\ref{fig:GalaxyMap}).   Due to the large-scale structure of the Galactic $\vec{B}$-field, the high-energy photon flux (produced via photon-ALP oscillations) has large-scale features on the gamma-ray sky.  As such, we will focus on two high-energy gamma-ray telescopes which can observe large patches of the sky: HAWC and LHAASO. 

The HAWC detector is a ground based high-energy gamma-ray observatory\,\cite{Abeysekara:2017hyn,Abeysekara:2017mjj,Abeysekara:2017hyn}.  It detects gamma-rays in the energy range $\gtrsim$ 100 GeV.  High-energy gamma-rays convert to electron-positron pairs in our atmosphere.  Charged leptons are detected via their Cherenkov radiation in water tanks.  The shower morphology enables HAWC to discriminate between gamma-rays and cosmic-rays.

LHAASO is a next-generation air shower instrument designed to detect gamma-rays and cosmic-rays via electronic, muonic, and Cherenkov radiation.  It will measure both cosmic-rays and gamma-rays in the energy range 10$^{12}$ eV -- 10$^{17}$ eV and 10$^{11}$ eV -- 10$^{15}$ eV respectively\,\cite{DiSciascio:2016rgi,Vernetto:2016gro,Vernetto:2015urm,Liu:2016kro,Cui:2014bda}.

\begin{figure}
       \includegraphics[width=0.48\textwidth]{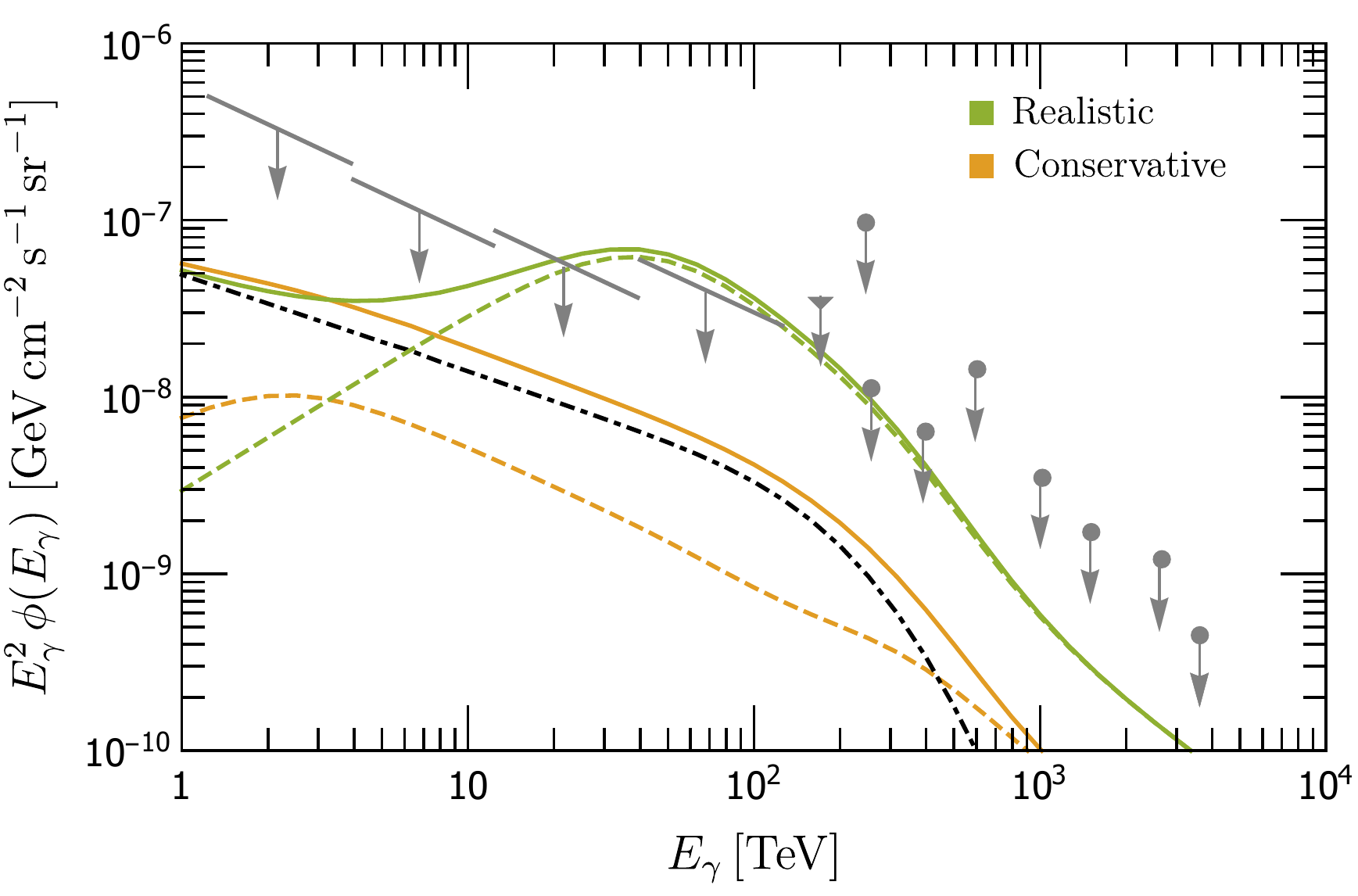}
  \caption{Gamma-ray spectrum for the direction $l=0^\circ$, $b=8^\circ$ for $g_{a\gamma}=6\times 10^{-11} \, \mm{GeV}^{-1}$ and $m_a = 10 \, \mm{neV}$. The components of the full flux (solid lines) are the ALP-induced photon-flux (dashed) and the photon background from $p-p$ interactions in the MW (black, dot-dashed). Limits are in gray: HAWC\,\cite{Abeysekara:2017wzt} (lines), GAMMA~\cite{Martirosov:2009ni} (triangle), and KASKADE-Grande~\cite{Apel:2017ocm} (dots).}\label{fig:GammaSpectrum}
\end{figure}

In Fig.~\ref{fig:GammaSpectrum}, we show example spectra of high-energy photons that will be incident on Earth in the direction $l=0^\circ$, $b=8^\circ$ for $g_{a\gamma}=6\times 10^{-11}\, \mm{GeV}^{-1}$ and $m_a =10\, \mm{neV}$. The solid lines are total photon fluxes from the converted diffuse ALP-flux (dashed) and the Galactic gamma-ray background from $p-p$ interactions (dot-dashed). Green (yellow) curves show the result for the realistic (conservative) scenario. We show as gray lines the limit on the northern Fermi bubbles by HAWC~\cite{Abeysekara:2017wzt}. We present it here since the morphology of this signal (an ``ALP bubble'') is very similar to the Fermi bubbles (see Fig.~\ref{fig:GalaxyMap}).    We also include GAMMA~\cite{Martirosov:2009ni} limits, and KASCADE-Grande limits~\cite{Apel:2017ocm} on the isotropic diffuse gamma-ray flux.  The signal morphology in this case is not isotropic and we expect an on-off technique to yield much tighter constraints. This specific morphology is unique in the high-energy gamma-ray sky, and this information can be used to mitigate backgrounds, or distinguish from local sources.

We showed our main results in Fig.~\ref{fig:MainResult}, which highlight the sensitivity of these gamma-ray detectors to ALPs. The HAWC result has been obtained by adjusting the upper limit in Ref.~\cite{Abeysekara:2017wzt} to ten years observation time with an effective uptime of $70\%$ and a $\sqrt{\rm time}$ scaling\,\footnote{private communication with Hugo Alberto Ayala Solares}. We compare that limit to our predicted photon flux in the field of view of HAWC. The estimate for HAWC is rather coarse and can be improved with real data.

For LHAASO, we model the detector using Ref.~\cite{Cui:2014bda}. For details of the sensitivity study see the supplemental materials.

{\itshape Discussion.---}For reasonable assumptions about the $\vec{B}$-field and source evolution, especially our detection estimates are encouraging. Note that we included only a fraction of the available $\vec{B}$-fields, and the photon-flux could be modified if the photons propagated through cluster or jet $\vec{B}$-fields, or the cosmic web~\cite{Montanino:2017ara}.  For robust bounds, the search for a diffuse ALP-flux suffers from these model dependencies and current unknowns like the sources' structure and evolution.  These uncertainties will certainly decrease in the future, and especially our knowledge about the type of sources will improve, giving us a more robust handle on these model parameters.

For the gamma-ray background and its spectral distortion through ALPs, we tested several dust models inspired by Refs.\,\cite{Misiriotis:2006qq,Ahlers:2013xia}. This background tends to obscure the ALP-induced photon flux in the galactic disc but was never important at higher latitudes. Our result is insensitive to the exact dust model.

Further improvement is expected on the experimental side. A telescope on the southern hemisphere with performance comparable to LHAASO will improve the test statistics by a factor ten. CTA might be able to detect the ALP bubble with the envisaged survey of the inner Galaxy~\cite{Acharya:2017ttl}.

The existence of ALPs will also impact Fermi-LAT's diffuse MeV--GeV background. Up to $33\%$ of the photon energy associated with IceCube's neutrinos can be converted into ALPs, which reduces the amount of cascaded photons seen by Fermi. A more complete analysis would require an extension of the density matrix (Eqn.~\ref{eq:propagation}) to include such cascades. Similarly, strong absorption in the sources (e.g. Ref.~\cite{Mannheim:1998wp}) requires a more detailed solution of the transport equation. 

{\itshape Conclusions.---}Searching for ALPs is one of the major endeavors in the low-energy frontier of beyond Standard Model physics.  There have been plenty of laboratory-based and astrophysics searches for ALPs, but an unambiguous signal has not yet been detected.  In this work, we propose a {\it new} technique to search for a diffuse high-energy ALP-flux.

The discovery of high-energy astrophysical neutrinos by IceCube guarantees the presence of a high-energy photon background.  Due to strong attenuation in the radiation fields of the sources, intergalactic space, and the Milky Way, these high-energy photons are exponentially suppressed on their way to Earth.  This expectation will dramatically change if ALPs are present.  The high-energy photons can convert to ALPs in the sources, which travel unimpeded through intergalactic space.  These ALPs will then convert back to high-energy photons in the MW's magnetic field.  The resulting photons can be potentially detected by present and near future high-energy gamma-ray experiments: HAWC, CTA, and LHAASO.  Our work is a unique blend utilizing the knowledge of high-energy astrophysical neutrinos and photon-ALP oscillations.  

Our results (Fig.~\ref{fig:MainResult}) show that this technique allows HAWC and LHAASO to detect or constrain a diffuse ALP-flux. In particular, these experiments will be able to probe parts of the dark matter parameter space, and regions that next-generation laboratory-based experiments, ALPS-II and IAXO, will be sensitive to.

Since the diffuse ALP-flux is emitted by an accumulation of (unknown) sources, we made some generic assumptions about the sources' mean environmental parameters and their redshift evolution. Our search strategy is guaranteed to deliver improved results as we discover  more about these sources.  This technique has fundamental advantages over other astrophysical searches for ALPs.  The present techniques either look at spectral features of individual sources or a certain source class at low redshift (typically $z \lesssim 1$), where detailed knowledge of the source is required. We are looking at a collection of different classes of sources situated all over the Universe.  Hence, the attenuation of these high-energy gamma-rays is much more severe and it will be very difficult for in-source astrophysics to explain an indication of ALPs in these searches. The proposed signal is a smoking-gun for ALPs.

\section*{Acknowledgments} 
We thank Markus Ahlers, John Beacom, Joseph Conlon, Sovan Chakraborty, Songzhan Chen, Alexander Friedland, Ignacio Izaguirre, Alexander Kartavtsev, Joachim Kopp, Alessandro Mirizzi, Cristina Popescu, Giuseppe Di Sciascio, Subir Sarkar, Georg Raffelt, Javier Redondo, Pasquale Serpico, G\"unter Sigl, Hugo Alberto Ayala Solares, and Silvia Vernetto for discussions. We are also especially grateful to Matt Kistler for his comments and discussions. H.V. is supported by the Department of Energy,  under
contract DE-AC02-76SF00515.  R.L. is supported by German Research  Foundation  (DFG)  under  Grant  Nos.  EXC-1098, KO 4820/1-1, FOR 2239, and from the European Research Council (ERC) under the European Union's Horizon 2020 research and innovation programme  (grant  agreement  No.  637506,  ``$\nu$Directions") awarded to Joachim Kopp.  M.M. is a Feodor-Lynen Fellow and acknowledges support of the Alexander von
Humboldt Foundation.

\bibliographystyle{kp}
\bibliography{Bibliography/references}	
\newpage

\onecolumngrid
\begin{center}
 \textbf{\large Supplemental Materials}
\end{center}
\vspace*{0.2cm}
\twocolumngrid

\section{LHAASO sensitivity analysis}
In this section we describe how we arrived at the stated sensitivity for the LHAASO experiment in Fig.~\ref{fig:MainResult}. We closely follow the results given in Ref.~\cite{Cui:2014bda}, which allows us to obtain sensible results without performing a complete detector simulation.

We test the ALP hypothesis ($H_1$) against a background-only hypothesis ($H_0$). The background hypothesis assumes that all observed photons either stem from cosmic-ray interactions with the atmosphere that are miss-identified as photons, or from Galactic gamma-rays produced by $p-p$ interactions. Hypothesis $H_1$ predicts gamma-rays from the diffuse extragalactic ALP flux as well as distortions in the Galactic gamma-ray spectrum in addition to the atmospheric background.

 The number of events detected by LHAASO for each process $i$ in the energy bin $E_j$ and solid angle $\Omega_k$ is given by
\begin{align*}
N_{i}(E_j,\Omega_k)=\int\D \Omega'_k\D E'_j A_{\mm{eff},i}(E'_j)J_i(E'_j,\Omega'_k)\epsilon_i(E'_j)\Delta t\,.
\end{align*}
Here, $A_{\mm{eff},i}$ is the effective area for gamma-rays or cosmic-rays taken from Ref.~\cite{Cui:2014bda}. $\epsilon_i$ is the efficiency to identify a photon for an incident gamma-ray, which is taken to be one in the following, or to miss-identify a cosmic-ray as a photon. The miss-identification rates are taken from Ref.~\cite{Vernetto:2015urm}. $\Delta t$ is the exposure time, which we will take to be ten years with an efficiency of $25\%$. $J_i$ is the incident differential flux. For the photon-flux induced by the extragalactic diffuse ALP-flux, we proceed as described in the main text. For cosmic-rays, this flux is taken from Ref.~\cite{Gaisser:2013bla}. For Galactic gamma-rays, we compute the flux by assuming that the spectrum Ref.~\cite{Gaisser:2013bla} holds in the entire galaxy and $p-p$ interactions with Galactic gas produce high-energy photons. For the gas models, we used variations of Refs.~\cite{Misiriotis:2006qq, Ahlers:2013xia}. The MW photon-flux is absorbed using the radiation fields by Ref.~\cite{Vernetto:2016alq}, and, under $H_1$, distorted by photon-ALP oscillations.

As a test statistic we use the Gaussian maximum-likelihood estimator with Poissonian standard deviation:
\begin{equation}
\chi^2 = \sum_{j,k}\frac{\left(N_{1,j,k}-N_{0,j,k}\right)^2}{\sqrt{N_{0,j,k}}}\,,
\end{equation}
where $N_{1,j,k}$ ($N_{0,j,k}$) is the number count in energy bin $j$ and direction $k$ under hypothesis $H_1$ ($H_0$). The energy binning is taken from Ref.~\cite{Cui:2014bda}, which we also used to cross-check that are assumptions yield comparable sensitivities to the Crab spectrum. Some deviations were found in the lowest-energy bin, which we ignored in this analysis.

For the sum over directions in the sky, we split LHAASO's field of view into patches of $2^\circ$ in declination and right ascension. We used LHAASO's proposed location at $29.35^\circ$ latitude with a field of view of $\pm40^\circ$ in declination (see Fig. 2 in Ref.~\cite{Liu:2016kro}).

 We assume that this statistic is $\chi^2$ distributed with two degrees of freedom. Since the systematics of LHAASO are not known to us, we perform the likelihood test with statistical errors only.

\end{document}